\documentclass[preprint,twocolumn,proceedings]{rmaa}

\SetYear{2002}
\SetConfTitle{Galaxy Evolution: Theory and Observations}

\title{Structure of Bright 2MASS Galaxies: 2D Fits to the $K_s$-band Surface Brightness Profiles}

\author{D.~H.~McIntosh\altaffilmark{1}, 
A.~H.~Maller\altaffilmark{1}, 
N.~Katz\altaffilmark{1},
M.~D.~Weinberg\altaffilmark{1}}

\altaffiltext{1}{University of Massachusetts, Amherst, Massachusetts 01003}

\suppressfulladdresses

\listofauthors{D. H. McIntosh, A. H. Maller, N. Katz, M. D. Weinberg}
\indexauthor{McIntosh, D. H.}

\addkeyword{galaxies: fundamental parameters}
\addkeyword{galaxies: structure}
\addkeyword{large-scale structure of universe}
\addkeyword{surveys}

\begin{document}
\maketitle 

\boldabstract{
The unprecedented sky coverage and photometric uniformity of
2MASS
provides a rich resource for obtaining a detailed
understanding of the galaxies populating our local ($z<0.1$) Universe.  
A full
characterization of the
physical structure of nearby galaxies is essential for theoretical and
observational studies of galaxy evolution and structure formation.
We have begun a quantified description of the internal structure and morphology
of 10,000 bright ($10\leq K_s\leq 11$) 2MASS galaxies
through multi-component model fits to the 2D surface brightness
profiles.
}

The current understanding of galaxy properties has been strongly biased
by the small fraction of hot, young stars that dominate observations in
optical wavelength surveys.  Our goal is a comprehensive survey of
local galaxy structure at near-infrared (NIR) 
wavelengths where the light best reflects
the total stellar mass and the effects of dust are minimized.
The 2MASS Extended Source Catalog contains over $3\times10^5$
galaxies at $K_s=13.1$
($10\sigma$), with $98\%$ reliability and $90\%$ completeness
(Jarrett et al. 2000, AJ, 119, 2498).
Outside the Galactic plane there are $\sim10^4$ 2MASS galaxies of
sufficient brightness ($10\leq K_s\leq 11$) and size ($\sim1\arcmin$
isophotal diameters) to be readily fit using GIM2D 
(Simard et al. 2002, astro-ph/0205025).

We have selected GIM2D because it is well-tested, it accounts 
for seeing through PSF
convolution, and it provides a variety of model profiles including an
$r^{1/n}$ bulge plus exponential disk.
For each galaxy, GIM2D produces model and residual images
(Fig. 1), plus quantitative
measures of internal structure (sizes, ellipticities, surface brightnesses) and
morphology (residual structure asymmetries, light concentration, relative
bulge and disk light contribution, bar strength).

\begin{figure}[!t]
  \includegraphics[width=\columnwidth]{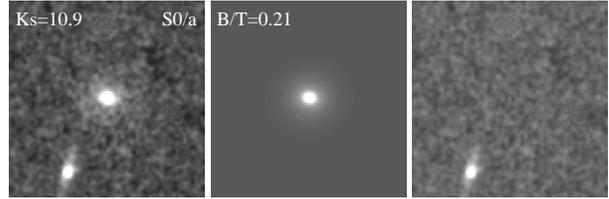}
  \caption{Example 2MASS $K_s$-band galaxy image (left), 
best-fit GIM2D $r^{1/4}$ bulge plus exponential disk model (middle),
and residual (model-subtracted) image (right).}
\end{figure}

We have tested the reliability of our fitting method
through a Monte Carlo procedure using a preliminary subset of
77 galaxies (39 E's and 38 S's)
drawn randomly from RC3 in the range $7<K_S<12.5$.  The background sky
uncertainty is the
largest source of error in parameters derived from profile fitting.
From the best-fit model of each galaxy, we constructed 25
realizations, artificially placed them in different
regions of blank 2MASS sky, and fit these to obtain the distribution of GIM2D
derived parameter values.  
For galaxies brighter than $K_s=11$, the typical parameter errors due to sky
are $\leq10\%$ (Fig. 2).  
We will apply this procedure to {\it all} profile 
fits in our study, providing the community a reliable catalog of
structural parameters with formal errors.

\begin{figure}[!t]
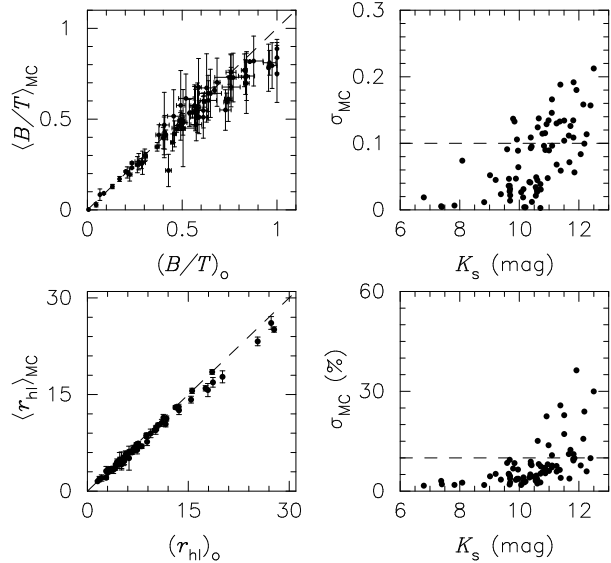

  \includegraphics[width=\columnwidth]{mcintosh_fig2.ps}
  \includegraphics[width=\columnwidth]{mcintosh_fig3.ps}
  \caption{Bulge-to-total ratio $B/T$ and half-light
radius $r_{\rm hl}$ (asecs)
error analysis for 77 galaxies.
Left: parameter
value from GIM2D fits to 2MASS galaxy image plotted against mean
from 25 Monte Carlo realizations with different random backgrounds.
Right: standard deviation of recovered
distribution as a function of
magnitude.}
\end{figure}





\end{document}